\documentclass[a4paper,10pt,twoside]{cpc-hepnp}
\usepackage{upgreek,fancyhdr}
\usepackage{multicol}
\usepackage{graphicx}
\usepackage{booktabs}
\usepackage{amssymb,bm,mathrsfs,bbm,amscd}
\usepackage[tbtags]{amsmath}
\usepackage{lastpage}
\usepackage{lineno}
\usepackage{subfigure}
\usepackage{amsmath}
\usepackage{overpic}

\begin{document}
%\linenumbers
%\begin{CJK*}{GB}{gbsn}
%\begin{CJK*}{GBK}{song}

\fancyhead[c]{\small Chinese Physics C~~~Vol. 41, No. 4 (2017) 046002}
\fancyfoot[C]{\small 010201-\thepage}

\footnotetext[0]{Received 20 September 2016}

\title{Study of $\mathrm{CdMoO_4}$ crystal for a neutrinoless double beta decay experiment with $\mathrm{{}^{116}Cd}$ and $\mathrm{{}^{100}Mo}$ nuclides\thanks{Supported by National Natural Science
Foundation of China (11275199) }}

\author{%
      Ming-xuan Xue $^{1,2;1)}$\email{xuemx@mail.ustc.edu.cn}%
\quad Yun-long Zhang $^{1,2;2)}$\email{ylzhang@ustc.edu.cn}%
\quad Hai-ping Peng $^{1,2}$\\
\quad Zi-zong Xu $^{1,2}$
\quad Xiao-lian Wang $^{1,2}$
}
\maketitle

\address{%
$^1$ State Key Laboratory of Particle Detection and Electronics (IHEP-USTC), Hefei 230026, China\\
$^2$ University of Science and Technology of China, Hefei 230026,  China\\
}

\begin{abstract}
The scintillation properties of a $\mathrm{CdMoO_4}$ crystal have been investigated experimentally. The fluorescence yields and decay times measured from 22 K to 300 K demonstrate that $\mathrm{CdMoO_4}$ crystal is a good candidate for an absorber for a bolometer readout, for both heat and scintillation signals. The results from Monte Carlo studies taking the backgrounds from $\mathrm{2\nu2\beta}$ of $\mathrm{{}_{42}^{100}Mo}$ ($\mathrm{{}_{48}^{116}Cd}$) and internal trace nuclides $\mathrm{{}^{214}Bi}$ and $\mathrm{{}^{208}Tl}$ into account show that the expected sensitivity of $\mathrm{CdMoO_4}$ bolometer for neutrinoless double beta decay experiment with an exposure of 100 $\mathrm{{kg}\cdot{years}}$ is one order of magnitude higher than those of the current sets of the $\mathrm{\lim{T^{0\nu\beta\beta}_{1/2}}}$ of $\mathrm{{}_{42}^{100}Mo}$ and $\mathrm{{}_{48}^{116}Cd}$.
\end{abstract}



\footnotetext[0]{\hspace*{-3mm}\raisebox{0.3ex}{$\scriptstyle\copyright$}2017
Chinese Physical Society and the Institute of High Energy Physics
of the Chinese Academy of Sciences and the Institute
of Modern Physics of the Chinese Academy of Sciences and IOP Publishing Ltd}%

\section{Introduction}

Almost two decades ago, the discovery of neutrino oscillation, a major achievement of particle physics, indicated that neutrinos have a non-vanishing rest mass \cite{lab1,lab2,lab3}. However, although neutrino oscillation experiments have probed the differences between neutrino mass states, the absolute mass scale is still unknown. The important challenge is to determine whether neutrinos are Dirac or Majorana particles. Dirac neutrinos can obtain mass through the standard Higgs mechanism like other leptons in the Standard Model; Majorana neutrinos act as their own antiparticles and acquire mass through the see-saw mechanism \cite{lab4,lab5}.

A golden channel for answering both the questions of neutrino nature and neutrino mass is neutrinoless double beta decay ($0\nu\beta\beta$),(Z,A)$\rightarrow$(Z+2,A)+$2e^{-}$. In most $0\nu\beta\beta$ experiments the signature of the decay is rather poor; it is possible for background events to mimic all the observables of $0\nu\beta\beta$ process. It is commonly accepted by the ``$\beta\beta$ community'' that the discovery of $0\nu\beta\beta$ would require that the decay shows up in more than one experiment for more than one nuclide.

Considering the detectors suitable for a rare-event search, a detector integrated with target nuclides, called as ``detector = source'' approach, is the first choice \cite{lab6,lab7,lab8,lab9}. The cryogenic phonon-scintillation detector is a promising detector to search for the $0\nu\beta\beta$ process. If the bolometer is a scintillating crystal, the heat signal can be combined with the light signal. The simultaneous detection of heat and light by bolometers has many advantages: the bolometric technique offers good energy resolution performance and excellent particle discrimination capability. The $\mathrm{CdMoO_4}$ crystal has several properties that make it a promising detector-source crystal for the bolometer: two interesting target nuclides, $\mathrm{{}^{100}Mo}$ and $\mathrm{{}^{116}Cd}$, are integrated into the crystal with fair natural abundance (Table~\ref{tab1}); and the Q-values of both nuclides (Table~\ref{tab1}) are well above the $\gamma$ (2615 keV) line of $\mathrm{{}^{208}Tl}$ trace nuclide.

\begin{center}
\tabcaption{ \label{tab1}  Properties of $\mathrm{{}^{100}Mo}$ and $\mathrm{{}^{116}Cd}$.}
\footnotesize
\begin{tabular*}{170mm}{@{\extracolsep{\fill}}ccccccc}
\toprule Parent isotope & Isotopic abundance (\%) \cite{lab10}& Q value (keV) \cite{lab11}& $\mathrm{T^{2\nu\beta\beta,exp}_{1/2}}$ (years) \cite{lab12}& $\mathrm{T^{0\nu\beta\beta}_{1/2}}$ (years) \cite{lab13,lab14}\\
\hline
$\mathrm{{}^{100}Mo}$ & 9.82 & 3034 & (7.1$\pm$0.4)$\times10^{18}$ & $>1.1\times10^{24}$\\
$\mathrm{{}^{116}Cd}$ & 7.49 & 2813 & (2.9$\pm$0.2)$\times10^{19}$ & $>1.7\times10^{23}$\\
\bottomrule
\end{tabular*}%
\end{center}

To explore the feasibility and capability of searching for $0\nu\beta\beta$ process on both $\mathrm{{}^{100}Mo}$ and $\mathrm{{}^{116}Cd}$ nuclides using a heat-scintillation bolometer with $\mathrm{CdMoO_4}$ as a detector-source crystal, the rest of this paper is arranged as follows. Section 2 presents an  experimental study of the scintillation properties of $\mathrm{CdMoO_4}$ crystal. Section 3 presents an internal background study, and Section 4 gives an evaluation of the sensitivity to $\mathrm{T^{0\nu\beta\beta}_{1/2}}$ of $\mathrm{{}^{100}Mo}$ and $\mathrm{{}^{116}Cd}$. Section 5 gives our conclusion and discusses future prospects.

\section{Experimental study of scintillation properties of $\mathrm{CdMoO_4}$ crystal}
\subsection{Instrumentation}

Figure 1 shows the experimental setup for measuring the characteristics of the $\mathrm{CdMoO_4}$ crystal. A $\mathrm{5\times5\times1}$ $\mathrm{mm^3}$ sample of the scintillator was placed inside the cryostat, which had one optical window to allow a laser beam in to stimulate the crystal and out to collect the emission light. The crystal characteristics were measured under the excitation by a 355-nm light from an Opolette 355 LD OPO system (Opotek Inc., with pulse length 7 ns and pulse repetition rate 20 Hz). Via an HRD1 double-grating monochromator (Jobin-Yvon), the light was collected with a photomultiplier (Hamamatsu R928-type). The data were output to an EG\&G 7265 DSP lock-in amplifier, and then recorded by a computer. The decay-time curve data were recorded on a Tektronix TDS2024 oscilloscope, and then input into the computer. Figure 2 shows $\mathrm{CdMoO_4}$ crystals grown by Ningbo University.

%\begin{overpic}[width=0.9\textwidth]{Fig1a.eps}%
%\put(50,10){(a)}
%\end{overpic}
\begin{figure}[!htbp]
\begin{center}
\subfigure[]{
\includegraphics[width=12cm]{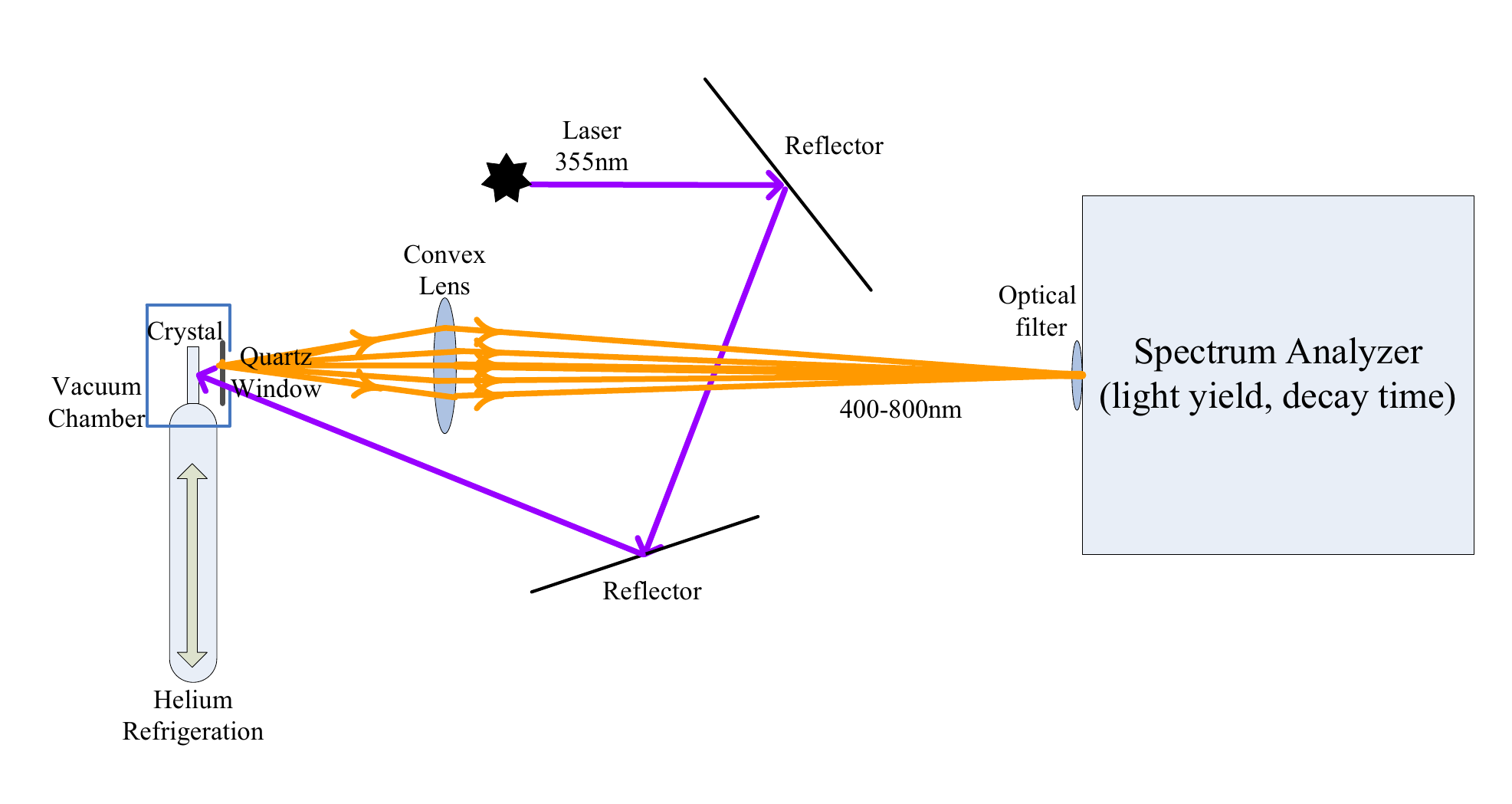}}
\subfigure[]{
\includegraphics[width=12cm]{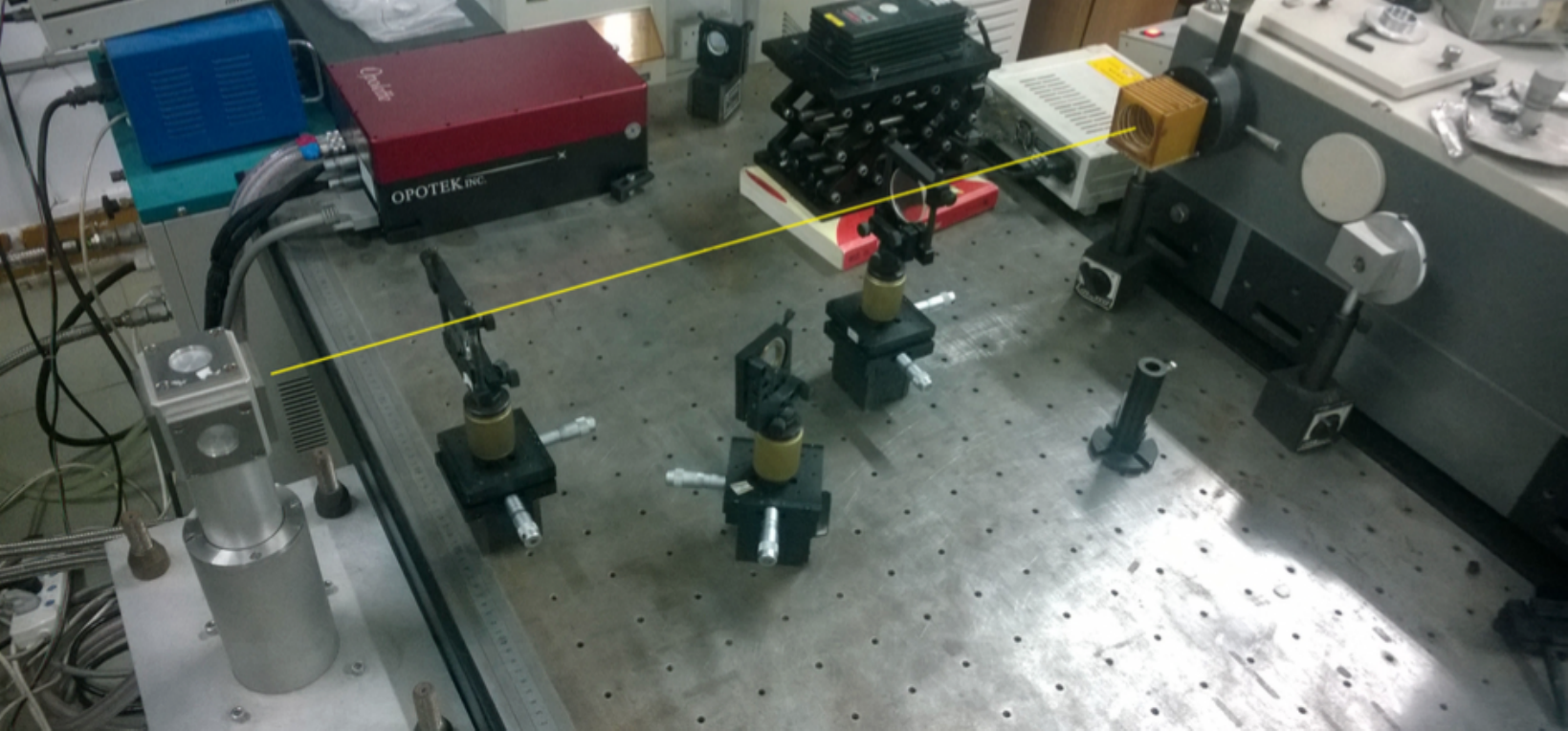}}
\figcaption{Experimental setup used to measure decay time and emission spectra. (a) Schematic diagram. (b) Photograph of physical setup.}
\end{center}
\end{figure}

\begin{center}
\includegraphics[width=6cm]{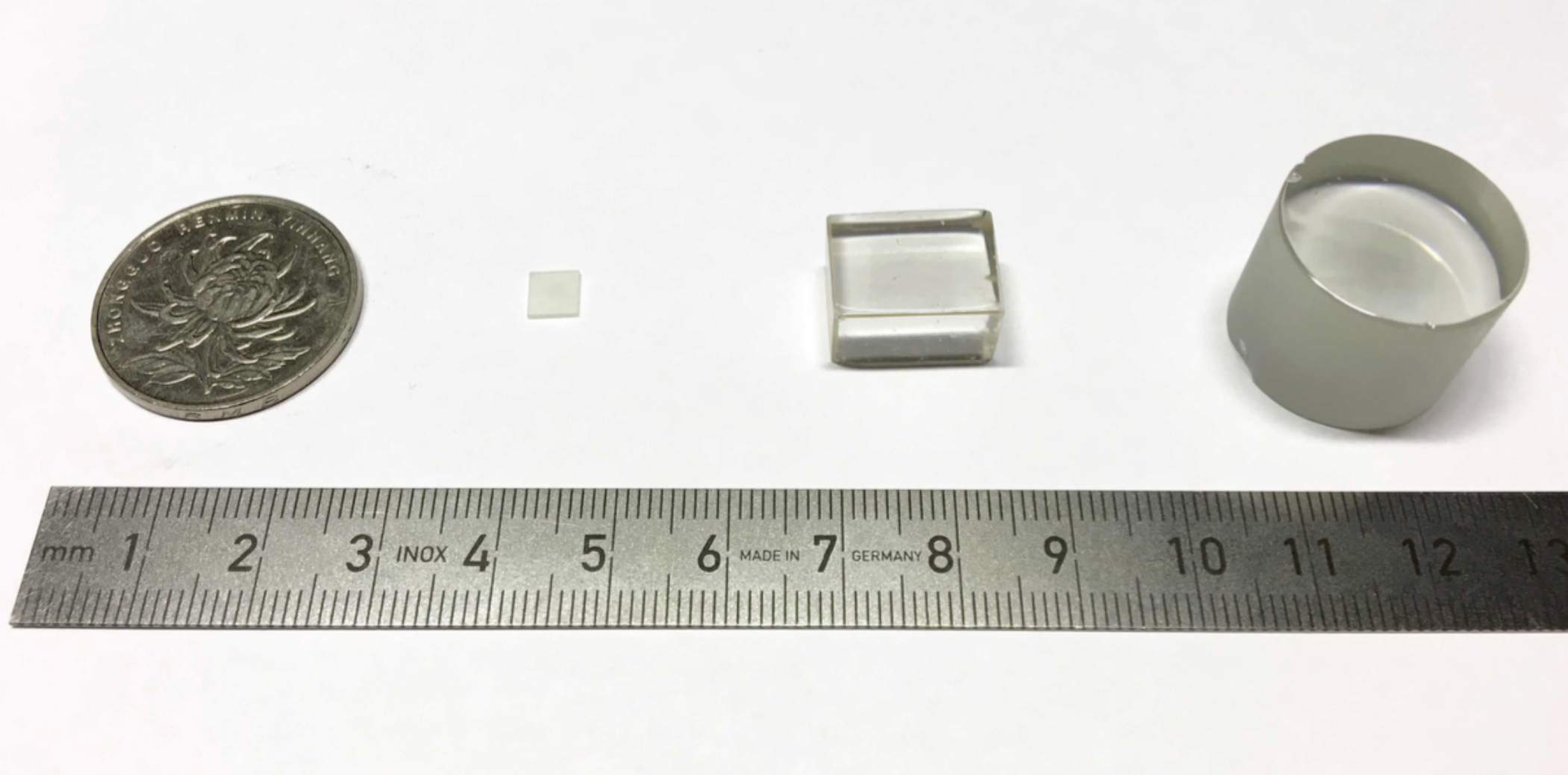}
\figcaption{$\mathrm{CdMoO_4}$ crystals of different shapes, produced by Ningbo University.}
\end{center}

\subsection{Scintillation properties}

Studies of temperature dependence of light yield and decay time give an opportunity to gain insight into the features of the scintillation process in the material. The use of $\mathrm{CdMoO_4}$ as a detector requires knowledge of several low-temperature characteristics, of which the luminescence properties are especially important. A $\mathrm{CdMoO_4}$ crystal excited with a laser beam of 355-nm wavelength exhibited broad emission bands that peaked at 551 nm (Fig. 3(a)). At room temperature, a $\mathrm{CdMoO_4}$ crystal emitted very faint light. With decreasing temperature, the light yield reached a maximum at approximately 150 K. Fig. 3(b) shows the measured dependence of the relative light yield of the $\mathrm{CdMoO_4}$ crystal scintillator on the temperature. According to cosmic ray experiments in the laboratory at room temperature, the light yield is about 10$\sim$20 phe/MeV.

\begin{figure}[!htbp]
\begin{center}
\subfigure[]{
\includegraphics[height=4cm]{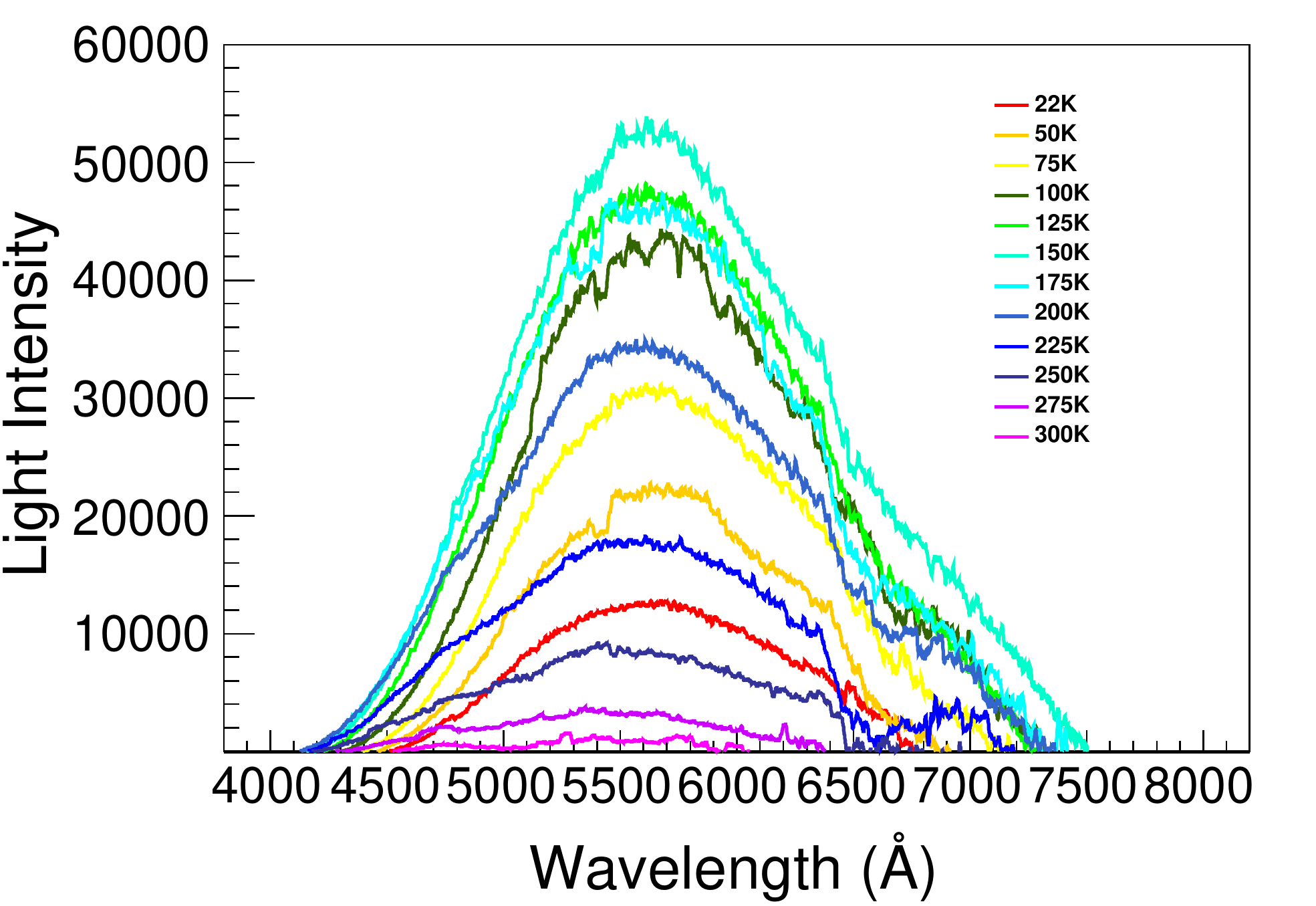}}
\subfigure[]{
\includegraphics[height=4.25cm]{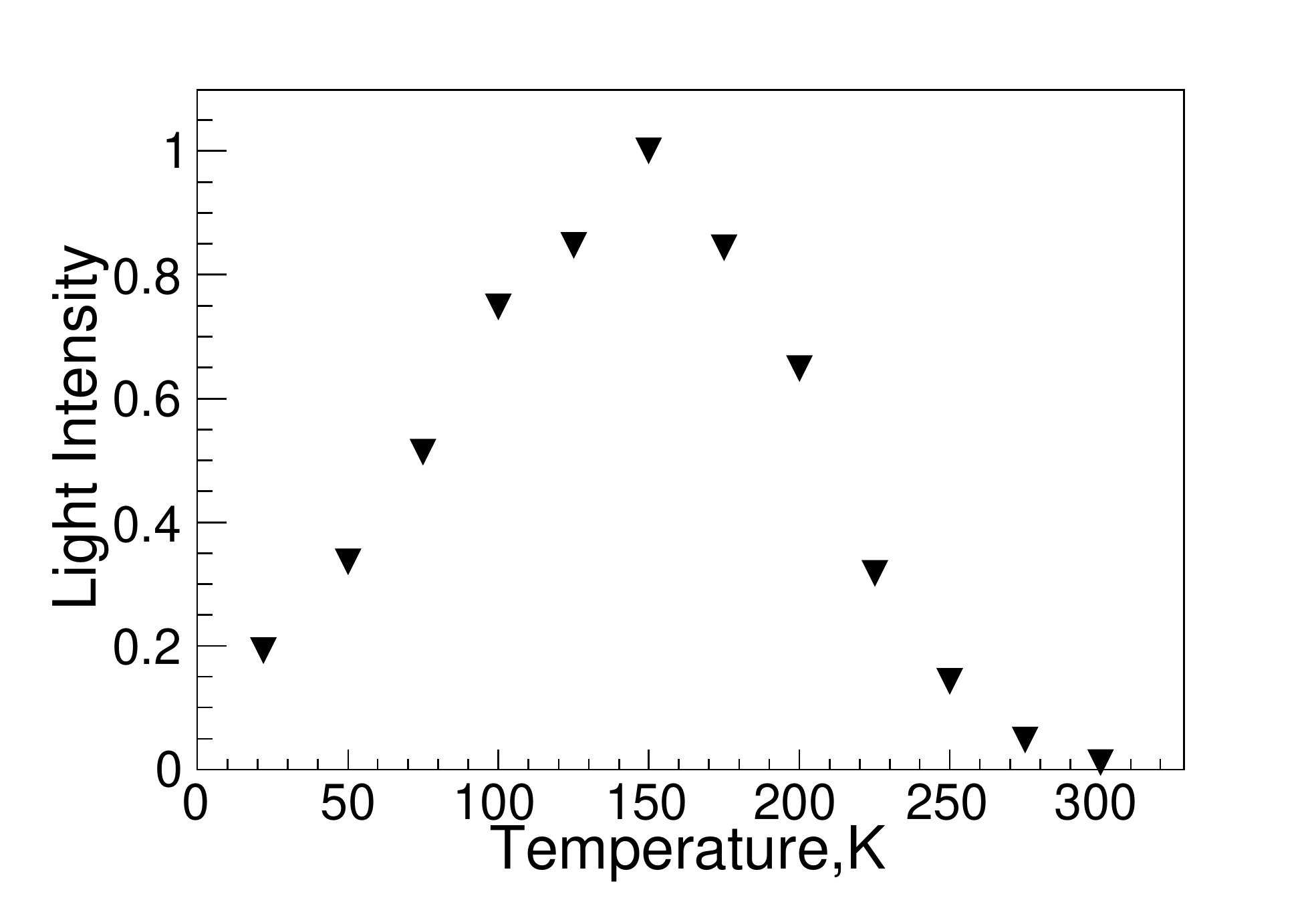}}
\figcaption{(color online)(a) Laser-induced emission spectra of a $\mathrm{CdMoO_4}$ crystal at different temperatures. (b) Temperature dependence of the luminescence intensity of a $\mathrm{CdMoO_4}$ crystal.}
\end{center}
\end{figure}

Fitting the luminescence pulse data, and the typical decay-time spectra of $\mathrm{CdMoO_4}$ measured at 22, 150, and 300 K are shown in Fig.4. The main decay-time constant was found to be 1.2 $\mathrm{\mu{s}}$ at room temperature (T=300 K); cooling to 22 K increased the scintillation decay-time constant to 170 $\mathrm{\mu{s}}$. Figure 5 displays the variation of the scintillation decay-time constant of $\mathrm{CdMoO_4}$ as a function of temperature. The temperature dependence of decay time for a $\mathrm{CdMoO_4}$ scintillator was qualitatively consistent with those of previous investigations of this class of materials \cite{lab15,lab16}.

\begin{figure}[!htbp]
\begin{center}
\subfigure[]{
\includegraphics[width=5.5cm]{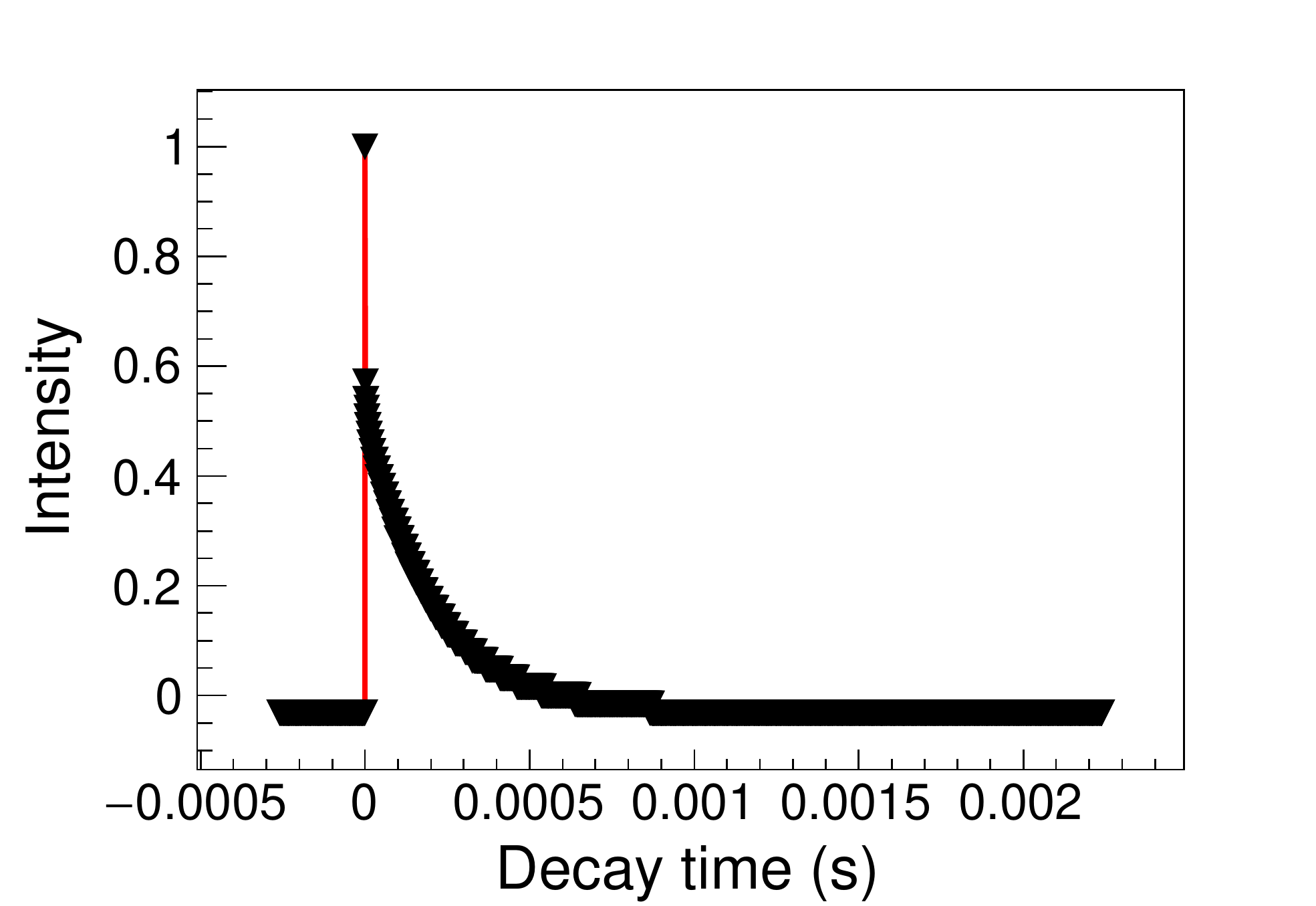}}
\subfigure[]{
\includegraphics[width=5.5cm]{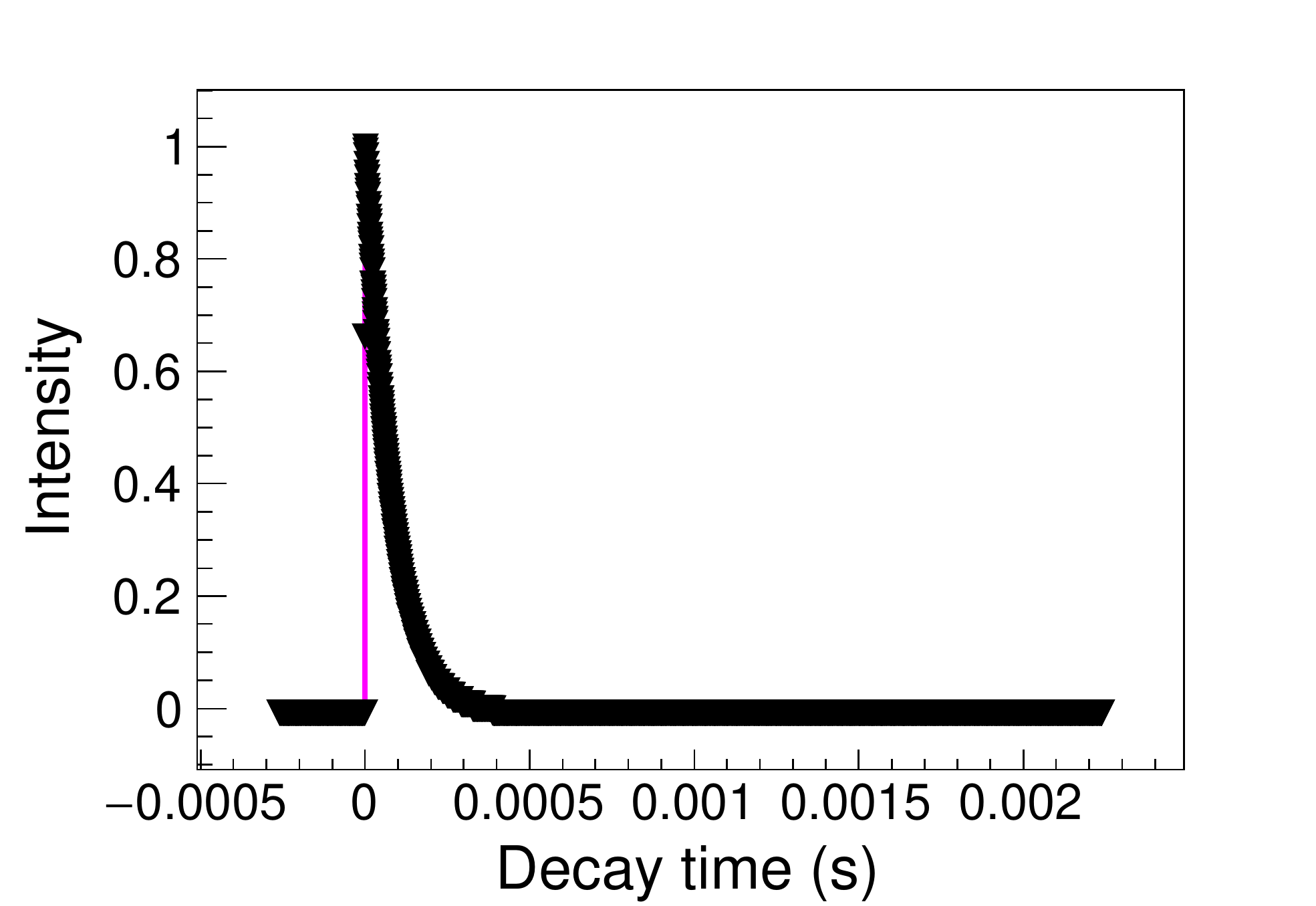}}
\subfigure[]{
\includegraphics[width=5.5cm]{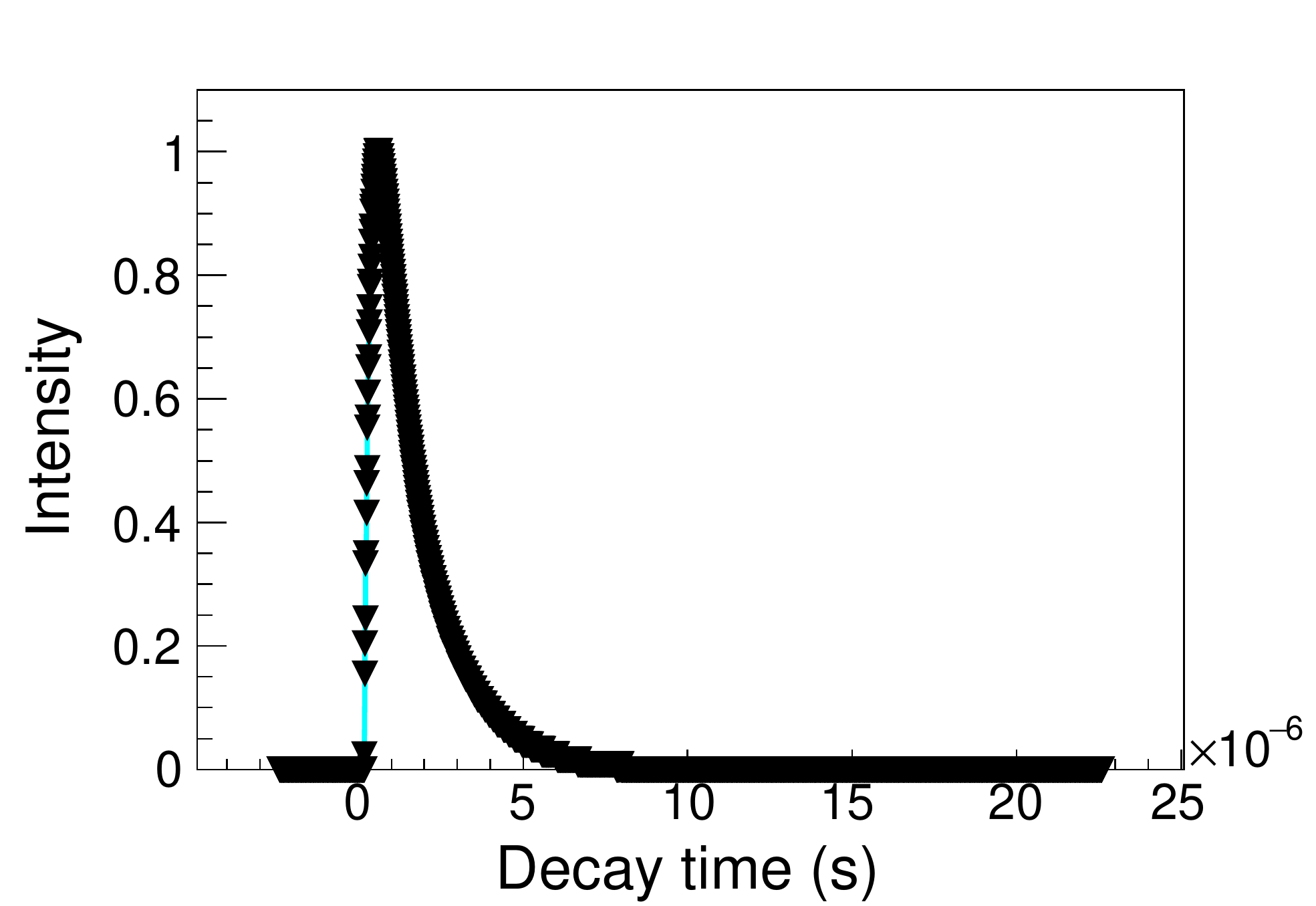}}
\figcaption{Decay-time spectrum of $\mathrm{CdMoO_4}$ at (a) 22 K, (b) 150 K and (c) 300 K.}
\end{center}
\end{figure}

\begin{figure}[!htbp]
\begin{center}
\includegraphics[width=6cm]{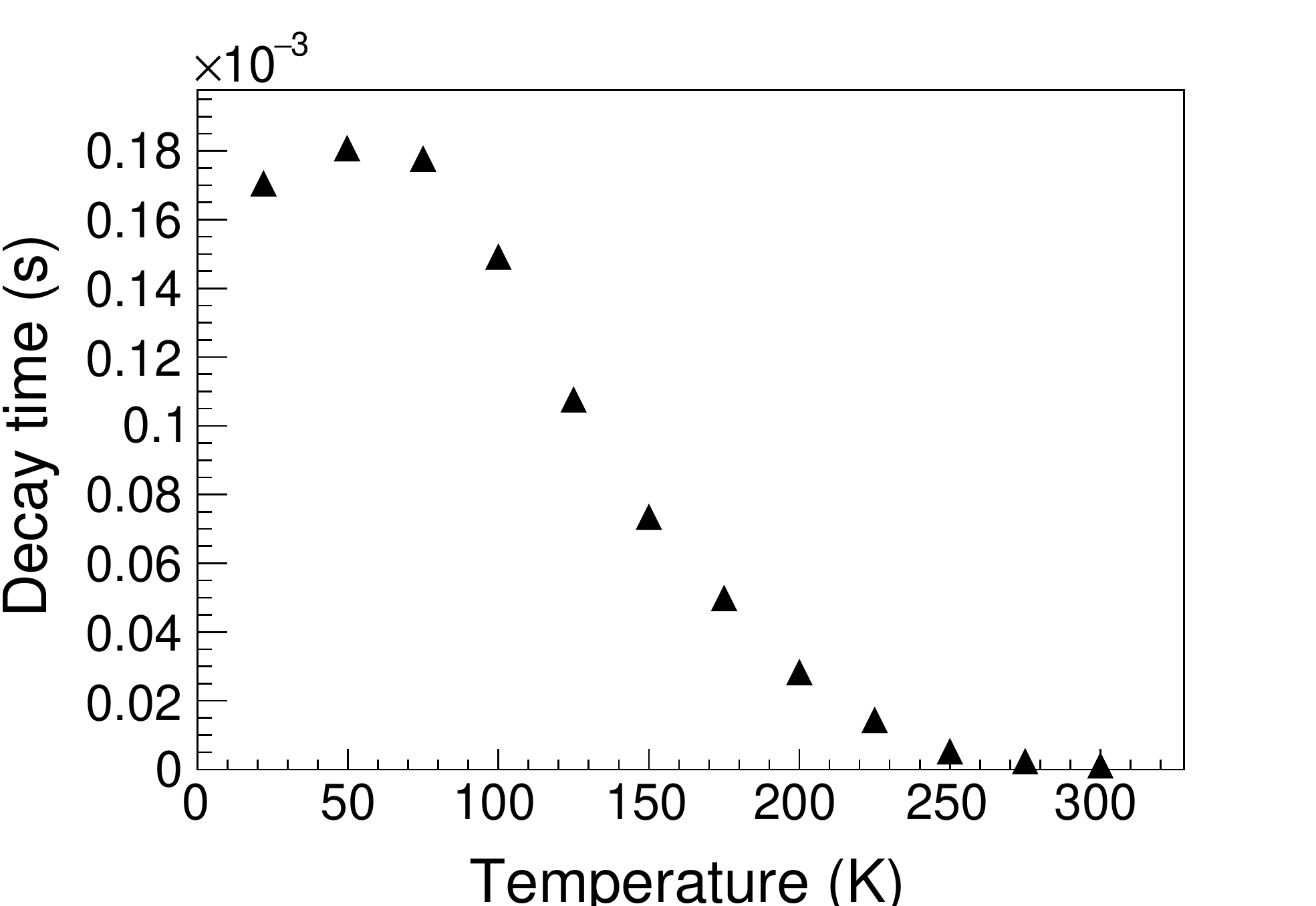}
\figcaption{Temperature dependence of the principal scintillation decay-time constant measured with a $\mathrm{CdMoO_4}$ detector under irradiation by a laser beam with a wavelength of 355 nm.}
\end{center}
\end{figure}

\section{Internal background study}

Events from the $0\nu\beta\beta$ process from $\mathrm{{}^{100}Mo}$ and $\mathrm{{}^{116}Cd}$ should appear in spectra around $3034\pm3\sigma_{E_1}$ and $2813\pm3\sigma_{E_2}$ respectively. This is called the Region Of Interest (ROI), where $\sigma_{E_i}$ is the square root of the variance @$E_i$. The background events falling in the ROI will directly limit the sensitivity of the measurement of $\mathrm{T^{0\nu\beta\beta}_{1/2}}$ and the significance of $0\nu\beta\beta$ signals. Two kinds of backgrounds were involved: 1) a continuous irremovable background from the $2\nu\beta\beta$ events of target nuclides $\mathrm{{}^{100}Mo}$ and $\mathrm{{}^{116}Cd}$, and 2) the background from the trace radio-nuclides $\mathrm{{}^{214}Bi}$ (in equilibrium with $\mathrm{{}^{226}Ra}$ from the $\mathrm{{}^{238}U}$ family) and $\mathrm{{}^{208}Tl}$ (in equilibrium with $\mathrm{{}^{228}Th}$ from $\mathrm{{}^{232}Th}$ family) \cite{lab17,lab18}. To estimate the influence of the backgrounds, 100\% enrichment in $\mathrm{{}^{100}Mo}$ and $\mathrm{{}^{116}Cd}$ was supposed, while the contributions from $\mathrm{{}^{214}Bi}$ and $\mathrm{{}^{208}Tl}$ were 0.1 mBq/kg activity \cite{lab18}, not considering shielding contamination.

\subsection{Backgrounds from $2\nu\beta\beta$ events of $\mathrm{{}^{100}Mo}$ and $\mathrm{{}^{116}Cd}$}

The continuous backgrounds from $2\nu\beta\beta$ events of $\mathrm{{}^{100}Mo}$ and $\mathrm{{}^{116}Cd}$ will hardly contaminate their own $0\nu\beta\beta$ peaks with good energy resolution. The key issue is the severity of the contamination of $0\nu\beta\beta$ peaks of $\mathrm{{}^{116}Cd}$ from $2\nu\beta\beta$ continuous spectrum of $\mathrm{{}^{100}Mo}$. GEANT4 simulations \cite{lab19} were used to model the shape of the energy spectra readout from the $\mathrm{CdMoO_4}$-bolometer. For the decay process of $\mathrm{{}^{100}Mo}$ and $\mathrm{{}^{116}Cd}$, the initial kinematics of the two emitted electrons were given by the DECAY0 event generator \cite{lab20}. In Fig. 6, energy resolution (using full width at half maximum (FWHM)) $R_{FWHM}$ of 1\%, 2\%, and 3\% were assumed. Information on the half-life span is given Table 1. In order to observe $0\nu\beta\beta$ signals of both $\mathrm{{}^{100}Mo}$ and $\mathrm{{}^{116}Cd}$ with proper significance, $R_{FWHM}$ should not be worse than 2\%@3 MeV while CUORE (Cryogenic Underground Observatory for Rare Events) has achieved the energy resolution goal of 5 keV FWHM at 2615 keV ($R_{FWHM}$=0.2\%) \cite{lab21}.

\begin{figure}[!htbp]
\begin{center}
\subfigure[]{
\includegraphics[width=5.5cm]{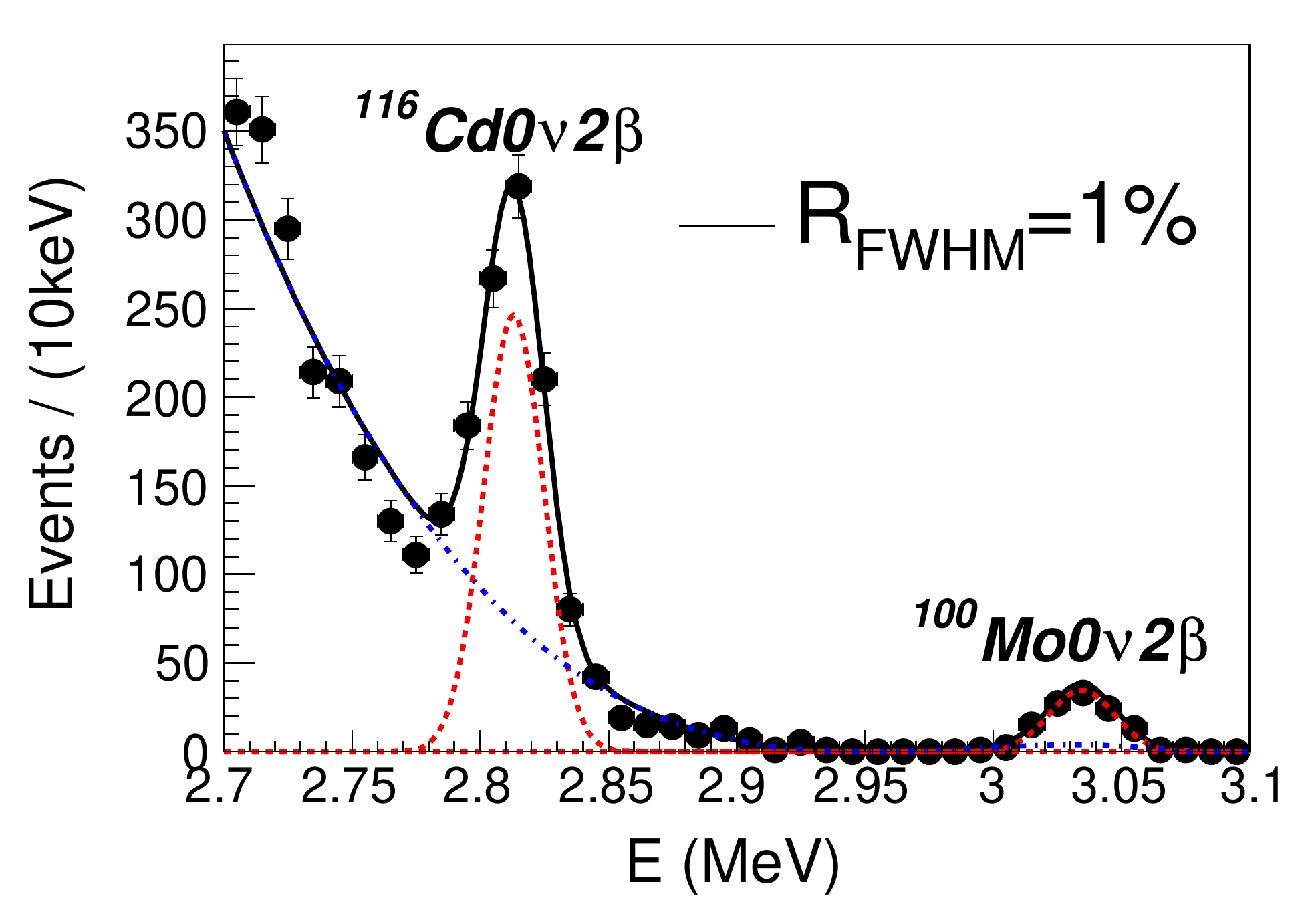}}
\subfigure[]{
\includegraphics[width=5.5cm]{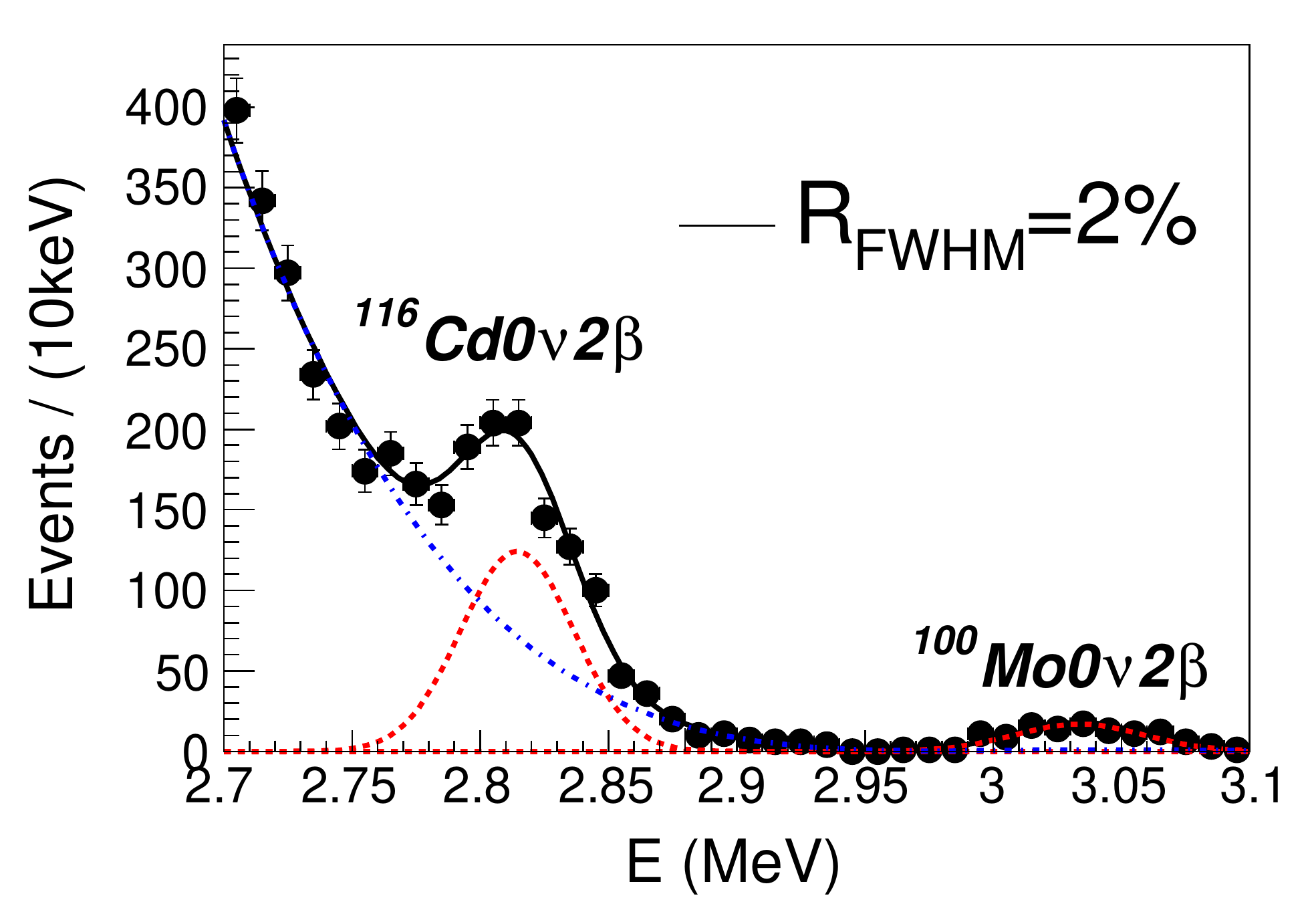}}
\subfigure[]{
\includegraphics[width=5.5cm]{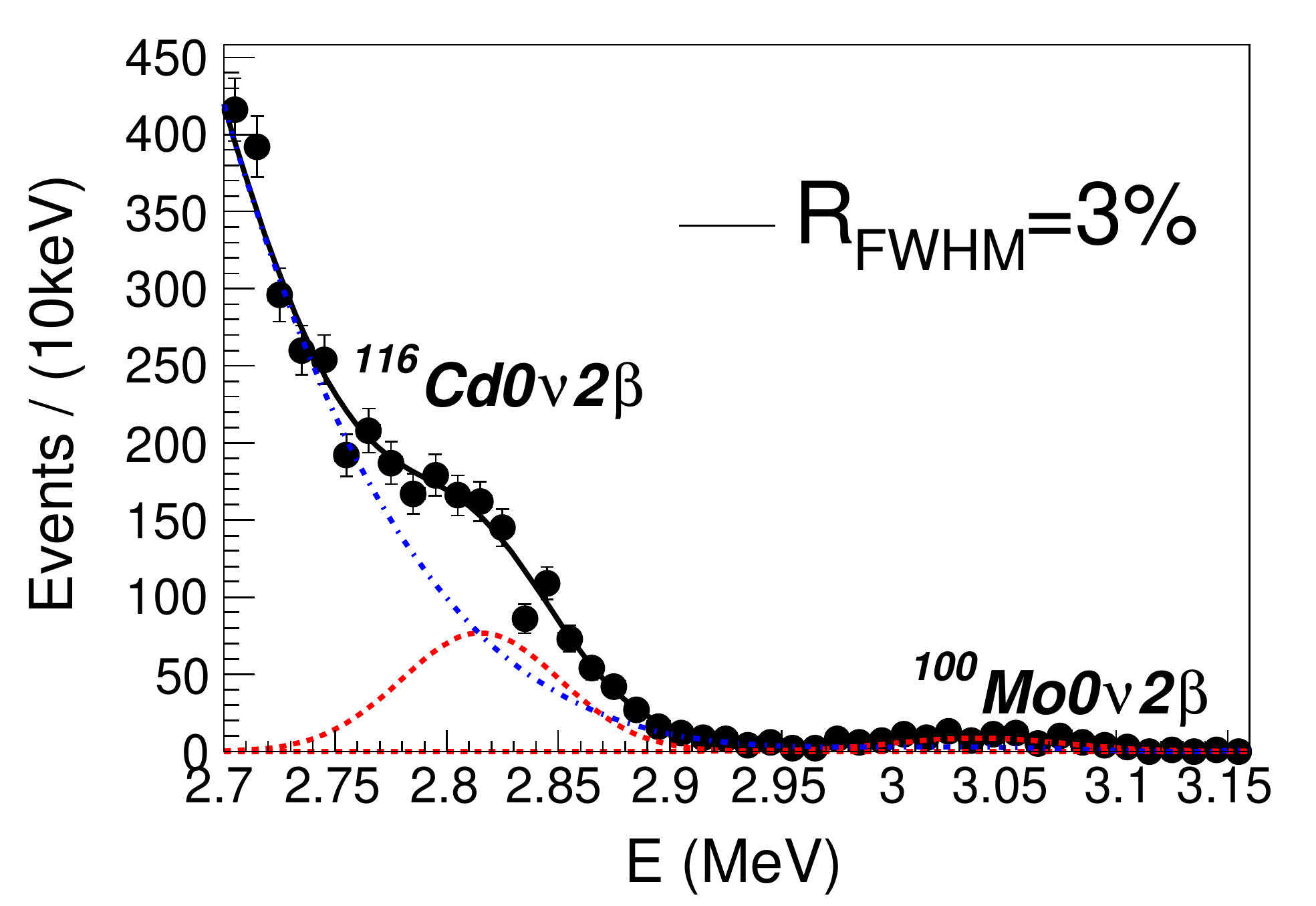}}
\figcaption{Different energy resolutions of the detector in GEANT4 simulations at the energies of $\mathrm{{}^{100}Mo}$ and $\mathrm{{}^{116}Cd}$ $0\nu\beta\beta$ decay ($R_{FWHM}$). (a) 1\%. (b) 2\%. (c) 3\%.}
\end{center}
\end{figure}

The preliminary results shown in Fig. 6(a) are convincing evidence that a heat-scintillation bolometer with $\mathrm{CdMoO_4}$ can be a promising design for searching for $0\nu\beta\beta$ events from $\mathrm{{}^{100}Mo}$ and $\mathrm{{}^{116}Cd}$ when the energy resolution is better than 1\%@3 MeV.

\subsection{Backgrounds from $\mathrm{{}^{214}Bi}$ and $\mathrm{{}^{208}Tl}$}

To estimate the internal backgrounds originating from $\mathrm{{}^{214}Bi}$ and $\mathrm{{}^{208}Tl}$ which are products of the $\mathrm{{}^{238}U}$ and $\mathrm{{}^{232}Th}$ decay chains respectively, we required a radiopure CsI(Tl) scintillation detector as an active shield. In GEANT4 simulations, a single detector module consists of a $\mathrm{5.5\times5.5\times5.5}$ $\mathrm{cm^3}$ crystal enriched in $\mathrm{{}^{100}Mo}$ and $\mathrm{{}^{116}Cd}$ to 100\%, surrounded by a CsI(Tl) scintillation detector of $\mathrm{50\times50\times50}$ $\mathrm{cm^3}$ that is just a phantom of the array of the $\mathrm{CdMoO_4}$ crystal. The mass of the $\mathrm{{}^{116}Cd{}^{100}MoO_4}$ detector is 1 kg with trace nuclides $\mathrm{{}^{214}Bi}$ and $\mathrm{{}^{208}Tl}$ of 0.1 mBq/kg.

Generally, a large internal contamination in the $\mathrm{{}^{238}U}$ chain could be worrisome due to one of its daughters; the decay chain $\mathrm{{}^{214}_{83}Bi\xrightarrow{\beta, Q=3272\ keV}{}^{214}_{84}Po\xrightarrow{\alpha, Q=7800\ keV}{}^{210}_{82}Pb}$ is of concern. The time characteristic of such an event is that an electron is followed by an alpha in a time interval of 163 $\mathrm{\mu{s}}$ ($\mathrm{T_{1/2}}$ of $\mathrm{{}^{214}_{84}Po}$). A time-amplitude identification method \cite{lab17,lab18}, called the beta-alpha coincidence method, can be used to reject these types of background events; when an energy deposit in the range of a few keV to 3272 keV happens, a check is made to determine whether an approximately 7800 keV deposit follows in a time window of 1 ms. Using the beta-alpha coincidence method, this background contribution can be further suppressed, as shown in Fig. 7. For better visualization of the suppression, the data are presented on a logarithmic scale. Thus, we are able to discriminate out 95\% of $\mathrm{{}^{214}Bi}$ background.

\begin{center}
\includegraphics[width=6cm]{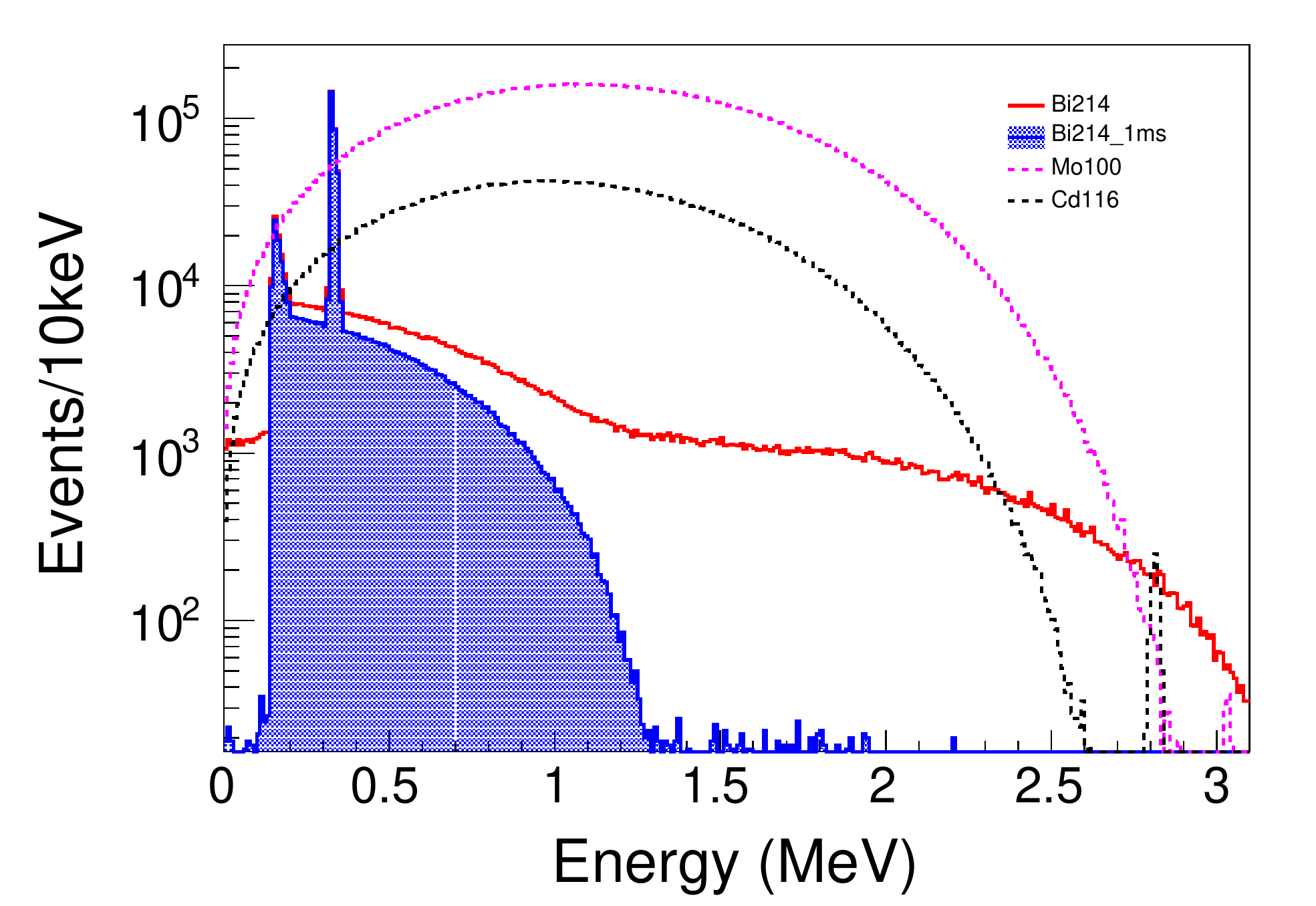}
\figcaption{(color online) Using a 1-ms time window to suppress the background from $\mathrm{{}^{214}Bi}$. The red line is when the coincidence method was not used. (0.1 mBq/kg activity)}
\end{center}

Another background source is $\mathrm{{}^{208}Tl}$ (from the $\mathrm{{}^{232}Th}$ family). The decay chain $\mathrm{{}^{208}_{81}Tl}\xrightarrow{\mathrm{\beta, Q_{gs}=5001\ keV}}\mathrm{{}^{208}_{82}Pb(e.s)}\xrightarrow{\mathrm{300 ps}}\mathrm{mult\gamma+{}^{208}_{82}Pb(g.s)}$ is taken. Considering the $\beta$ decay process of $\mathrm{{}^{208}Tl}$, it is accompanied by de-excitation of $\gamma$ rays from $\mathrm{{}^{208}Pb}$(e.s). The $\gamma$ rays in these kinds of background events will mostly escape from the target detector and be finally absorbed by the surrounding active shield detectors (the active shield detectors could be replaced by the array detector units around the one in which the event is being evaluated.). With a 4$\pi$ gamma veto system \cite{lab18} and ``one and only'' selection, most of the backgrounds associated with trace nuclide $\mathrm{{}^{208}Tl}$ will be suppressed, as shown in Fig. 8.

\begin{center}
\includegraphics[width=6cm]{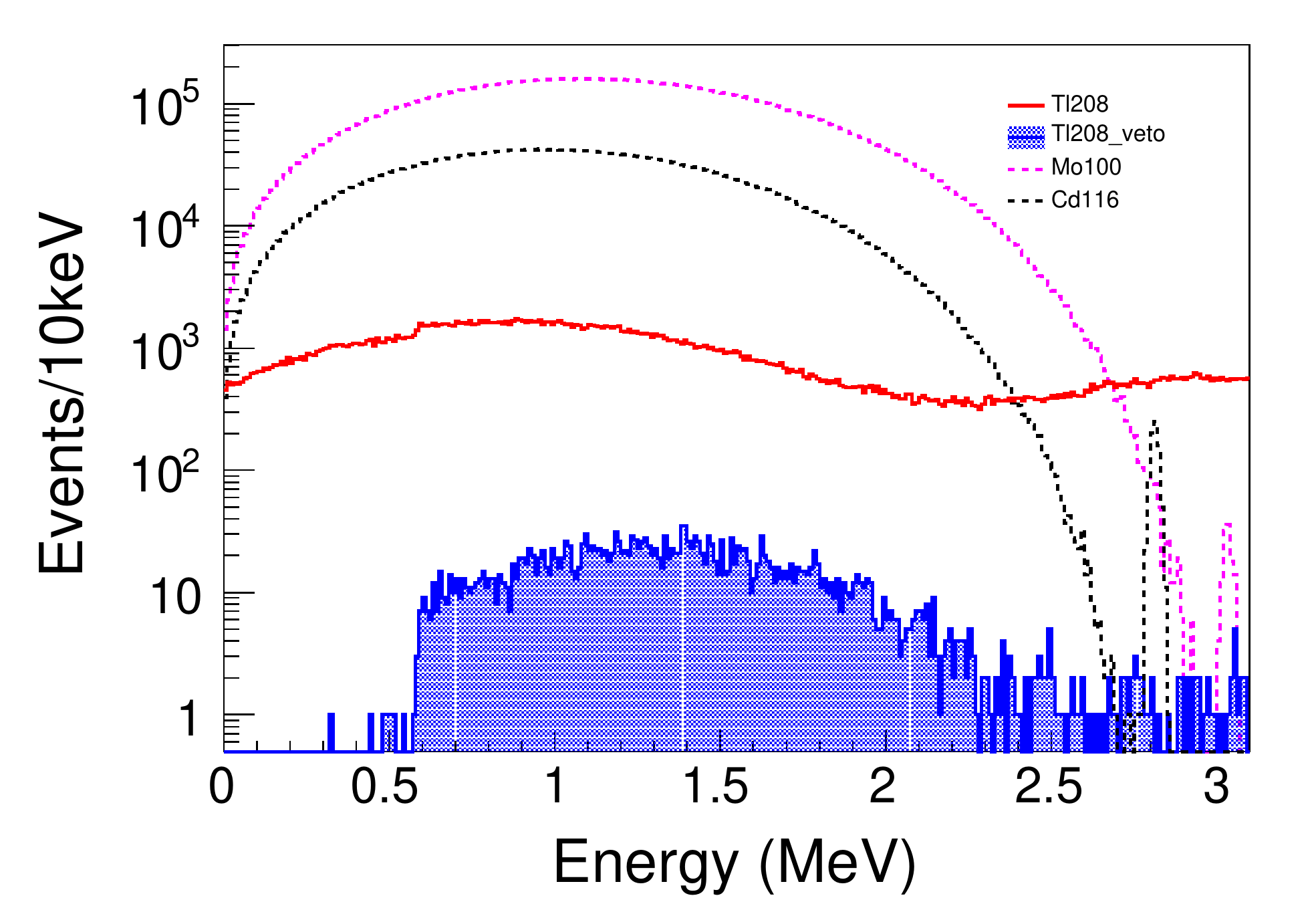}
\figcaption{(color online) Using the 4$\pi$ gamma veto system to decrease internal background from $\mathrm{{}^{208}Tl}$. The red line is when the anti-coincidence method was not used. (0.1 mBq/kg activity)}
\end{center}

\section{Evaluation of the sensitivity of the $\mathrm{CdMoO_4}$-bolometer for $\mathrm{\lim{T^{0\nu\beta\beta}_{1/2}}}$ of $\mathrm{{}^{100}Mo}$ and $\mathrm{{}^{116}Cd}$}

The scintillation properties and radioactive contamination of $\mathrm{CdMoO_4}$ crystals have been described above. To estimate the sensitivity of the $\mathrm{CdMoO_4}$-bolometer for limiting the $\mathrm{T^{0\nu\beta\beta}_{1/2}}$ of $\mathrm{{}^{100}Mo}$ and $\mathrm{{}^{116}Cd}$, the MC-data of the exposure of 100 $\mathrm{kg\cdot{years}}$ of $\mathrm{{}^{116}Cd{}^{100}MoO_4}$-bolometer was generated with $R_{FWHM}=1$\% of the bolometer, 0.1 mBq/kg of trace radioactivity of $\mathrm{{}^{214}Bi}$ and $\mathrm{{}^{208}Tl}$, and $\mathrm{T^{2\nu\beta\beta}_{1/2}}$ of $\mathrm{{}^{100}Mo}$ and $\mathrm{{}^{116}Cd}$ (Table 1). The signal and background spectra are shown in Fig. 6a, Fig. 7 and Fig. 8.

The sensitivity in terms of a half-life limit of $0\nu\beta\beta$ can be estimated using the known formula:

\begin{eqnarray}
\lim{T_{1/2}}\sim\ln{2}\cdot\varepsilon\cdot{N}\cdot{t}/\lim{S}(90\% C.L.)
\end{eqnarray}

where $\varepsilon$ is the detection efficiency, $N$ is the number of $\mathrm{{}^{100}Mo}$ ($\mathrm{{}^{116}Cd}$) nuclei in the scintillation crystal, $t$ is the measuring time, and $\lim{S}$ is the maximum number of $0\nu\beta\beta$ events which can be excluded with a given confidence level on Monte Carlo simulation background. The detection efficiency $\varepsilon$ was provided by GEANT4 simulation. A Bayesian approach \cite{lab21} estimated the upper limit of the $0\nu\beta\beta$ decay rate of $\mathrm{{}^{100}Mo}$ and $\mathrm{{}^{116}Cd}$. The predicted half-life sensitivity to $0\nu\beta\beta$ decay of the nuclides of $\mathrm{{}^{100}Mo}$ and $\mathrm{{}^{116}Cd}$ are $\mathrm{\lim{T^{0\nu\beta\beta}_{1/2}}=1.02\times{10^{25}}yr}$ and $\mathrm{\lim{T^{0\nu\beta\beta}_{1/2}}=3.68\times{10^{24}}yr}$ at 90\% C.L. respectively, almost one order of magnitude higher than those of the current sets (Table~\ref{tab1}).

\section{Conclusions and prospects}

The fluorescence properties measured show that $\mathrm{CdMoO_4}$ crystal is a suitable absorber for a heat-scintillation bolometer to search for neutrinoless double beta decay of $\mathrm{{}^{100}Mo}$ and $\mathrm{{}^{116}Cd}$. The Monte Carlo study provided convincing evidence that signals of $0\nu\beta\beta$ of $\mathrm{{}^{116}Cd}$ in the ROI would be higher than the background from the $2\nu\beta\beta$ events of $\mathrm{{}^{100}Mo}$. Using the beta-alpha coincidence and the 4$\pi$ gamma veto method, most of the background from $\mathrm{{}^{214}Bi}$ and $\mathrm{{}^{208}Tl}$ with 0.1 mBq/kg activity is suppressed in the ROI. New limits of $\mathrm{T^{0\nu\beta\beta}_{1/2}}$ of $\mathrm{{}^{100}Mo}$ and $\mathrm{{}^{116}Cd}$ are set with one order of magnitude improvement. A prototype of heat-scintillation bolometer using $\mathrm{CdMoO_4}$ is going to be fabricated. The trace radioactive nuclides in $\mathrm{CdMoO_4}$ and background identification will be extensively explored using this prototype bolometer.

\section{Acknownledgement}

\acknowledgments{The authors express their gratitude to Ningbo University for providing the $CdMoO_4$ crystals and the Department of Physics of the University of Science and Technology of China for providing the low temperature laboratory equipment. We would also like to thank Vladimir Tretyak for his generous help.}

\vspace{15mm}

\vspace{-1mm}
\centerline{\rule{80mm}{0.1pt}}

\clearpage
%\end{CJK*}
\end{document}